\documentclass[aps,pra,twocolumn,preprintnumbers]{revtex4}
\usepackage{amsmath}
\usepackage{amsmath,amssymb,bm,graphicx,subfigure,color}

\begin{document}

\title{Level crossing in the three-body problem for strongly interacting fermions
\\in a harmonic trap}
\author{J. P. Kestner, L.-M. Duan}
\affiliation{FOCUS center and MCTP, Department of Physics,
University of Michigan, Ann Arbor, MI 48109}

\begin{abstract}
We present a solution of the three-fermion problem in a harmonic
potential across a Feshbach resonance. We compare the spectrum with
that of the two-body problem and show that it is energetically
unfavorable for the three fermions to occupy one lattice site rather
than two. We also demonstrate the existence of an energy level
crossing in the ground state with a symmetry change of its wave
function, suggesting the possibility of a phase transition for the
corresponding many-body case.
\end{abstract}

\maketitle
\section{Introduction}
Ultracold atoms, tuned with Feshbach resonance, offer a great
opportunity to study strongly correlated many-body physics in a
controlled fashion. For such strongly interacting systems, in
general there is no well-controlled approximation method to solve
the many-body physics. Exact solution of few-body problems plays an
important role in understanding the corresponding many-body systems.
Few-body (three- or four-body) problems have been solved for
strongly interacting bosons or fermions in free space
\cite{Petrov03}, in a quasi-one-dimensional configuration
\cite{Mora05}, and in a three-dimensional (3D) harmonic trap for
bosons \cite{Stoll05} or fermions in the unitary limit
\cite{Werner06,Chang07}.

In this paper, we add to this sort of examples by exactly solving
the three-body problem for strongly interacting two-component
fermions in a 3D harmonic trap across resonance. This work has two
main motivations: Firstly, the situation considered here is relevant
for experiments where one loads strongly interacting two component
fermions into a deep 3D optical lattice \cite{lat}. For each site
that can be approximated with a harmonic potential, one could have
two identical fermions (spin-$\uparrow $) strongly interacting with
another distinct fermion (spin-$\downarrow $). The three-body
problem for equal mass fermions turns out to be very different from
the corresponding case for bosons. Instead of a hierarchy of bound
Efimov states for bosons \cite {Stoll05}, we show that there is
always a significant energy penalty for three fermions to occupy the
same lattice site ($\uparrow \uparrow \downarrow $) instead of two
($\uparrow \downarrow +\uparrow $). This result justifies an
important assumption made in the derivation of an effective
many-body Hamiltonian for this system \cite{Duan05}. Secondly, we
analyze the ground-state structure of the three-body problem and
show that as one scans the 3D scattering length, there is a level
crossing between the lowest-lying three-fermion energy eigenstates,
which have \textit{s}- or \textit{p}-wave symmetries respectively in
the limit as two atoms are contracted to form a dimer. This level
crossing with a symmetry change may correspond a quantum phase
transition in the many-body case where one has spin-polarized fermi
gas loaded into an optical lattice. The latter system with polarized
fermions has raised strong interest recently in both theory and
experiments \cite{exp06,theory06}.
\section{Methods}
The method here is based on manipulation of a Lippmann-Schwinger
equation for the wavefunction, with the formalism similar to the one
presented in Ref. \cite {Mora05}. As is standard, we separate the
center-of-mass degree of freedom via an orthogonal transformation of
variables, leaving the trapping potential diagonal in the new
coordinates. The Schr\"{o}dinger equation for the relative degrees
of freedom is then
\begin{multline}
\left[ -\frac{\hbar ^{2}}{m_{0}}\left( \nabla _{\mathbf{x}}^{2}+\nabla _{%
\mathbf{y}}^{2}\right) +\frac{1}{4}m_{0}\omega ^{2}\left( \mathbf{x}^{2}+%
\mathbf{y}^{2}\right) -E\right] \Psi \left( \mathbf{x},\mathbf{y}\right)  \\
=-\sum_{\pm }V\left( \mathbf{r}_{\pm }\right) \Psi \left( \mathbf{x},\mathbf{%
y}\right) \,,
\end{multline}
where $m_{0}$ is the atomic mass, $\omega $ is the trap frequency, $\mathbf{y%
}$ is the vector between the two $\uparrow $ fermions, $\sqrt{3}\mathbf{x}/2$
is a vector from the center of mass of the two $\uparrow $ fermions to the $%
\downarrow $ fermion, and $\mathbf{r}_{\pm }=\sqrt{3}\mathbf{x}/2\pm \mathbf{%
y}/2$ are the vectors from the $\downarrow $ fermion to each of the two $%
\uparrow $ fermions. We approximate the short-range interaction between
fermions with the usual zero-range pseudopotential \cite{Huang56} $V\left(
\mathbf{r}\right) =\frac{4\pi \hbar ^{2}a}{m_{0}}\delta \left( \mathbf{r}%
\right) \frac{\partial }{\partial r}\left( r\cdot \right) $, where
$a$ is the 3D s-wave scattering length tunable through the Feshbach
resonance.

The above contact interaction is equivalent to imposing boundary conditions $%
\Psi \left( \mathbf{x},\mathbf{y}\right) \simeq \mp \frac{f\left( \mathbf{r}%
_{\perp ,\pm }\right) }{4\pi \mathbf{r}_{\pm }}\left( 1-\frac{\mathbf{r}%
_{\pm }}{a}\right) $ for $\mathbf{r}_{\pm }\rightarrow 0$, where the $%
\mathbf{r}_{\perp ,\pm }=\mathbf{x}/2\mp \sqrt{3}\mathbf{y}/2$ are
proportional to the distances between the center of mass of an
$\uparrow \downarrow $ pair and an $\uparrow $ fermion. The overall
$\mp $ sign ensures the antisymmetry of the wavefunction upon
swapping the identical fermions. The undetermined function $f\left(
\mathbf{r}_{\perp ,\pm }\right) $, after a rescaling of the
argument, is the relative atom-dimer wavefunction that results when
two of the fermions form a tightly bound pair. Solving for this
asymptotic wavefunction $f \left( \mathbf{r}_{\perp ,\pm } \right) $
fully determines $\Psi \left( \mathbf{x}, \mathbf{y} \right) $.

Since $V\left( \mathbf{r}_{\pm }\right) $ only acts at $\mathbf{r}_{\pm }=0$%
, we use the asymptotic form of $\Psi $ when computing $V\left( \mathbf{r}%
_{\pm }\right) \Psi $. The formal solution can then be written as
\begin{multline}
\Psi \left( \mathbf{x},\mathbf{y}\right) =\int d\mathbf{x^{\prime }}d\mathbf{%
y^{\prime }} G_{E}^{\left( 2\right) }\left( \mathbf{x},\mathbf{y};\mathbf{%
x^{\prime }},\mathbf{y^{\prime }}\right)
\\
\times \sum_{\pm }\frac{\mp \hbar ^{2}f\left( \mathbf{r^{\prime
}}_{\perp ,\pm }\right) }{m_{0}}\delta \left( \mathbf{r^{\prime
}}_{\pm }\right) \label{eq:SE}
\end{multline}
where the two-particle Green's function is given by
\begin{equation}
G_{E}^{\left( 2\right) }\left( \mathbf{x},\mathbf{y};\mathbf{x^{\prime }},%
\mathbf{y^{\prime }}\right) =\sum_{\lambda _{1}\lambda _{2}}\frac{\psi
_{\lambda _{1}}\left( \mathbf{x}\right) \psi _{\lambda _{2}}\left( \mathbf{y}%
\right) \psi _{\lambda _{1}}^{\ast }\left( \mathbf{x^{\prime
}}\right) \psi _{\lambda _{2}}^{\ast }\left( \mathbf{y^{\prime
}}\right) }{E_{\lambda _{1}}+E_{\lambda _{2}}-E}  \label{eq:G2}
\end{equation}
The $\psi _{\lambda _{i}}\left( \mathbf{x}\right) $ are the single-particle
eigenfunctions with eigenenergies $E_{\lambda _{i}}$ for the reduced mass $%
m_{0}/2$. Here we use spherical coordinates for the three-dimensional
harmonic trap, so the quantum numbers are $\lambda =\left( n,l,m\right) $, $%
n=0,1,2,...$; $l=0,1,2,...$; $m=-l,-l+1,...,l-1,l$. The eigenenergies are $%
E_{\lambda }=\left( 2n+l+3/2\right) \hbar \omega $ and the
eigenfunctions are $\psi _{\lambda }\left( \mathbf{r}\right)
=R_{nl}\left( r\right) Y_{l}^{m}\left( \theta ,\phi \right) $, where
the $Y_{l}^{m}\left( \theta ,\phi \right) $ are the standard
spherical harmonics. The radial wavefunction is given by
\cite{Powell61}
\begin{equation}
R_{nl}\left( r\right) =\sqrt{\frac{2n!/d^{3}}{\left( n+l+1/2\right) !}}%
e^{-r^{2}/2d^{2}}\left( \frac{r}{d}\right) ^{l}L_{n}^{l+1/2}\left(
r^{2}/d^{2}\right)
\end{equation}
where $d=\sqrt{\frac{2\hbar }{m_{0}\omega }}$ is the length scale of
the trap and the $L_{n}^{k}\left( r\right) $ are associated Laguerre
polynomials.

We can make use of the invariance of $G_{E}^{\left( 2\right) }$ and the
integration measure under an orthogonal transformation of variables to
rewrite Eq. \eqref{eq:SE} in terms of $\mathbf{r}\equiv \mathbf{r}_{-}$ and $%
\mathbf{r}_{\perp }\equiv \mathbf{r}_{\perp ,-}$,
\begin{multline}
\Psi \left( \mathbf{r},\mathbf{r}_{\perp }\right) =\frac{\hbar ^{2}}{m_{0}}%
\int d\mathbf{r}_{\perp }^{\prime }f\left( \mathbf{r}_{\perp }^{\prime
}\right) \biggl[ G_{E}^{\left( 2\right) }\left( \mathbf{r},\mathbf{r}_{\perp };0,%
\mathbf{r}_{\perp }^{\prime }\right)   \label{eq:SE2} \\
-G_{E}^{\left( 2\right) }\left( \frac{\mathbf{r}}{2}+\frac{\sqrt{3}\mathbf{r}%
_{\perp }}{2},\frac{\sqrt{3}\mathbf{r}}{2}-\frac{\mathbf{r}_{\perp }}{2};0,%
\mathbf{r}_{\perp }^{\prime }\right) \biggr]
\end{multline}
We can also decompose the asymptotic atom-dimer wavefunction in terms of the
complete set of single-particle wavefunctions, $f\left( \mathbf{r}_{\perp
}\right) =\sum_{\lambda }f_{\lambda }\psi _{\lambda }\left( \mathbf{r}%
_{\perp }\right) $. Then Eq. \eqref{eq:SE2} becomes
\begin{multline}
\Psi \left( \mathbf{r},\mathbf{r}_{\perp }\right) =d^{2}\hbar \omega
\sum_{\lambda } \biggl[ G_{E-E_{\lambda }}\left( \mathbf{r},0\right)
\psi _{\lambda
}\left( \mathbf{r}_{\perp }\right)   \label{eq:SE3} \\
-G_{E-E_{\lambda }}\left( \frac{\mathbf{r}}{2}+\frac{\sqrt{3}\mathbf{r}%
_{\perp }}{2},0\right) \psi _{\lambda }\left( \frac{\sqrt{3}\mathbf{r}}{2}-%
\frac{\mathbf{r}_{\perp }}{2}\right) \biggr] f_{\lambda }
\end{multline}
where $G_{E}$ is the single-particle Green's function,
\begin{multline}
G_{E}\left( \mathbf{r},0\right) = \sum_{\lambda }\frac{\psi
_{\lambda }\left( \mathbf{r}\right) \psi _{\lambda }^{\ast }\left(
0\right) }{E_{\lambda }-E}
\\
= \frac{e^{-r^{2}/2d^{2}}}{2\pi ^{3/2}d^{3}\hbar \omega } \Gamma
\left( \frac{ \frac{3}{2} - E/\hbar \omega }{2} \right) U \left(
\frac{\frac{3}{2}-E/\hbar \omega }{2},\frac{3
}{2},\frac{r^{2}}{d^{2}} \right)
\end{multline}
\cite{Busch98}, and $U$ is the confluent hypergeometric function. Note that
the three-fermion wavefunction is fully determined by $f\left( \mathbf{r}%
_{\perp }\right) $ and that if we consider Eq. \eqref{eq:SE3} in the
limit as $\mathbf{r}\rightarrow 0$, we obtain a self-consistent
equation for $f\left( \mathbf{r}_{\perp }\right) $ by using the
boundary conditions. After some work, we obtain
\begin{equation}
\sum_{\lambda ^{\prime }}A_{\lambda \lambda ^{\prime }}f_{\lambda
^{\prime }}=\left[ \frac{d}{a}-2\frac{\Gamma \left(
\frac{3/2+E_{\lambda }/\hbar \omega -E/\hbar \omega }{2}\right)
}{\Gamma \left( \frac{1/2+E_{\lambda }/\hbar \omega -E/\hbar \omega
}{2}\right) }\right] f_{\lambda } \label{eq:eigen}
\end{equation}
where
\begin{equation}
A_{\lambda \lambda ^{\prime }} = \int \!\!\!\!
\frac{d\mathbf{r}_{\perp }}{4\pi d^{3}\hbar \omega } G_{E-E_{\lambda
^{\prime }}} \! \left( \! \frac{\sqrt{3} \mathbf{r}_{\perp }}{2}, 0
\! \right) \psi _{\lambda }^{\ast } \left( \mathbf{r}_{\perp
}\right) \psi _{\lambda ^{\prime }} \! \left( \frac{\mathbf{
-r}_{\perp }}{2}\right) \label{eq:A}
\end{equation}

\section{Results}
\begin{figure}[tbp]
\includegraphics[width=.85\columnwidth]{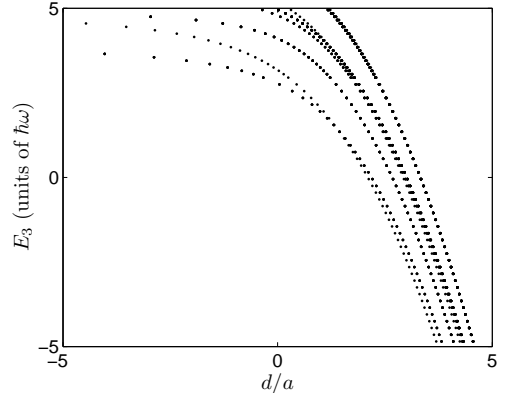}
\caption{Energy vs. inverse scattering length.} \label{fig:Evsa}
\end{figure}

\begin{figure}[tbp]
\includegraphics[width=.85\columnwidth]{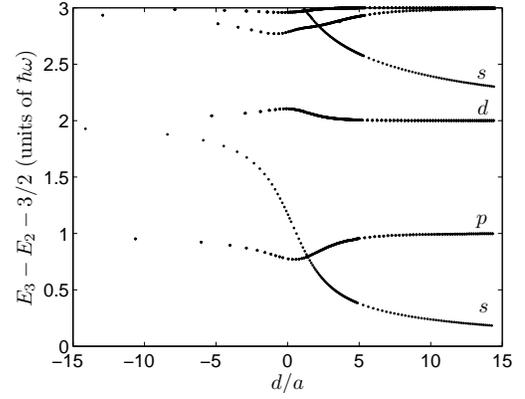}
\caption{Difference between three-fermion energy and two-fermion
energy plus one-fermion energy vs. inverse scattering length.}
\label{fig:Ediff}
\end{figure}
\begin{figure}[tbp]
\includegraphics[width=.25\columnwidth]{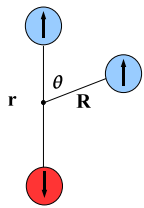}
\caption{(Color online) Relative coordinates for the three-fermion
problem. In terms of variables used in Eq. \eqref{eq:SE3}, $r =
\left| \mathbf{r} \right|$, $R = \protect\sqrt{3} \left|
\mathbf{r}_{\perp} \right| /2$.} \label{fig:vars}
\end{figure}

\begin{figure}[tbp]
\subfigure[$n=0$, $l=0$, $m=0$]
        {\includegraphics[width=.5\columnwidth]
        {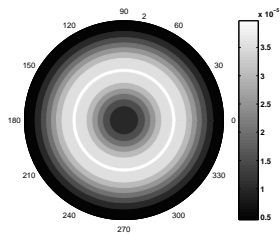}}
\subfigure[$n=0$, $l=1$, $m=\pm 1$]
        {\includegraphics[width=.5\columnwidth]
        {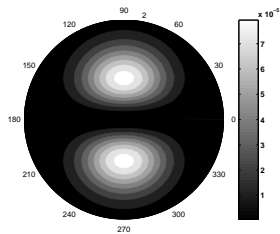}}
\subfigure[$n=0$, $l=1$, $m=0$]
        {\includegraphics[width=.5\columnwidth]
        {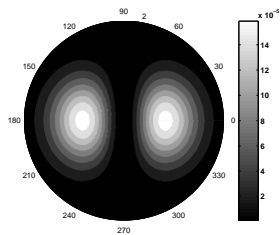}}
\caption{Contour plots of $r^2 \left| \Psi \left( r, R,
\protect\theta
\right) \right| ^2 _{r \rightarrow 0}$ as a function of $R$ and $\protect%
\theta$ for $r/d \simeq 0$ and $d/a \simeq 0$.} \label{fig:sym}
\end{figure}

\begin{figure}[tbp]
\subfigure[$n=0$, $l=0$, $m=0$]
        {\includegraphics[width=.5\columnwidth]
        {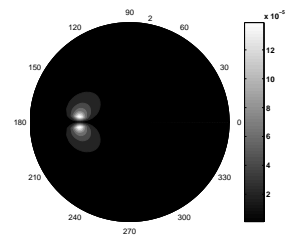}}
\subfigure[$n=0$, $l=1$, $m=\pm 1$]
        {\includegraphics[width=.5\columnwidth]
        {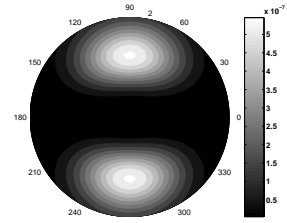}}
\subfigure[$n=0$, $l=1$, $m=0$]
        {\includegraphics[width=.5\columnwidth]
        {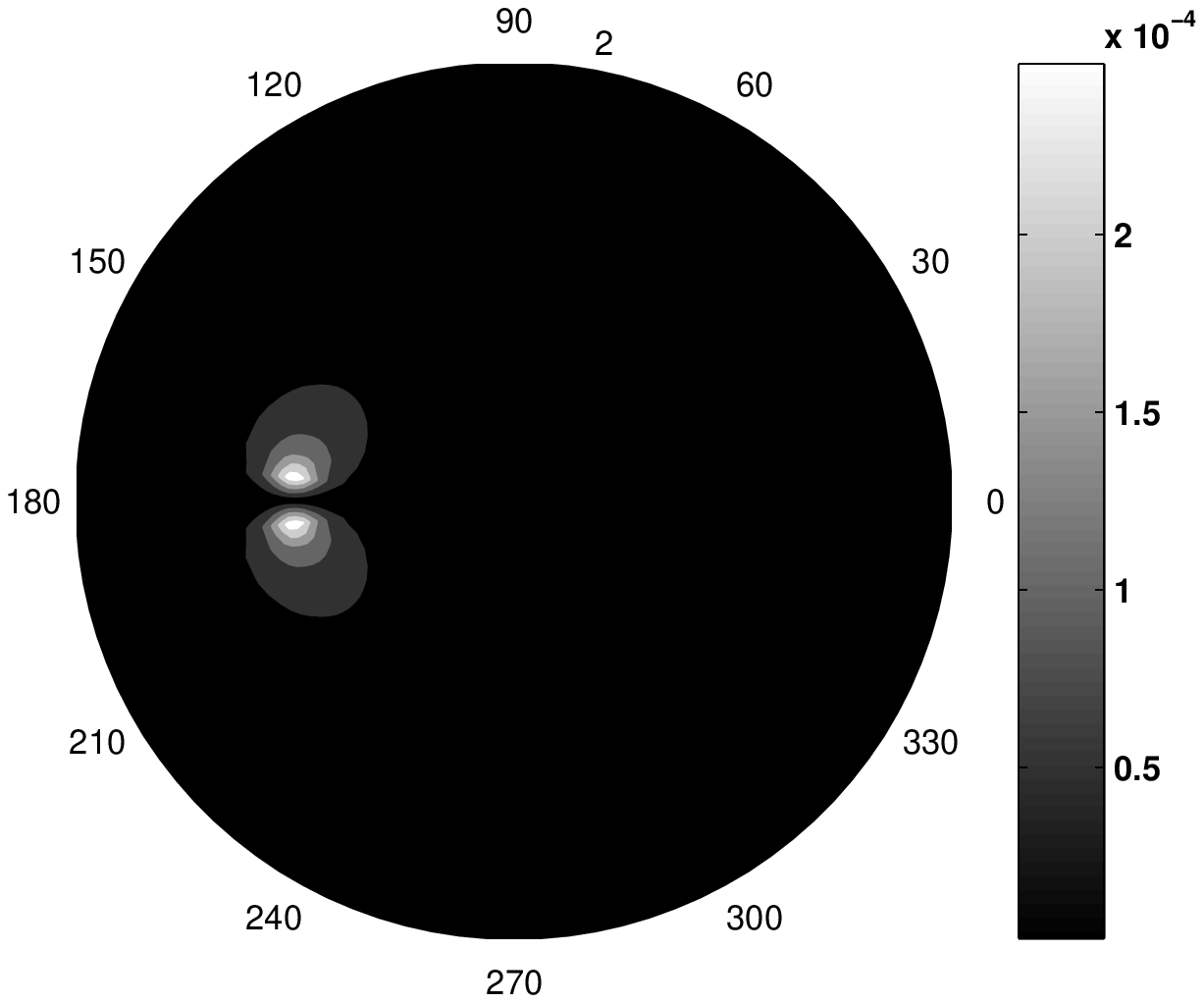}}
\caption{Contour plots of $r^2 R^2 \left| \sin \protect\theta \Psi
\left( r, R, \protect\theta \right) \right| ^2$ as a function of $R$
and $\protect \theta$ for $r/d=2$ and $d/a \simeq 0$.}
\label{fig:asym}
\end{figure}

\begin{figure*}[tbp]
\subfigure[Normalized probability density vs. $r/d$]
        {\includegraphics[width=2\columnwidth]
        {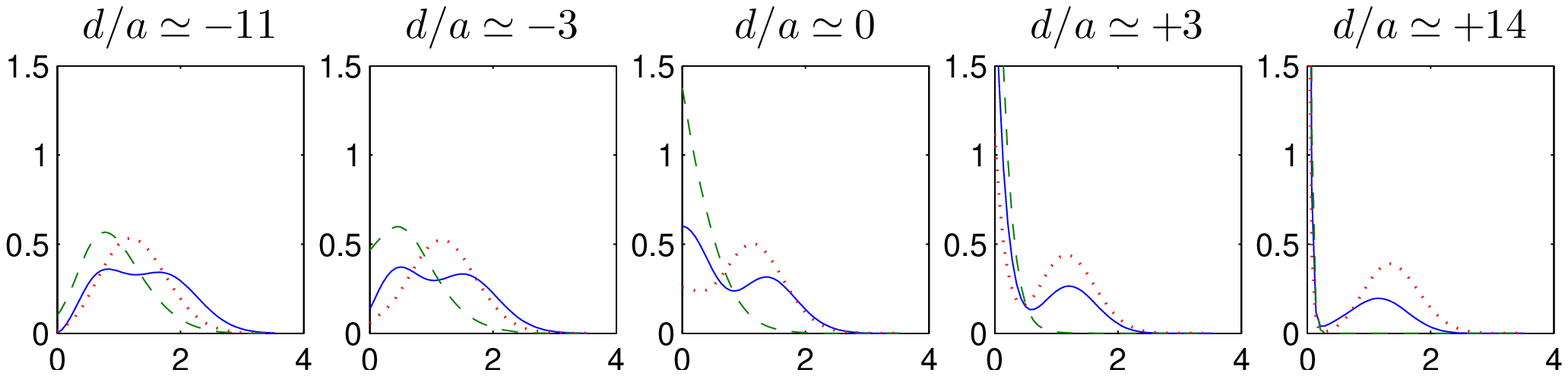}}
\subfigure[Normalized probability density vs. $R/d$]
        {\includegraphics[width=2\columnwidth]
        {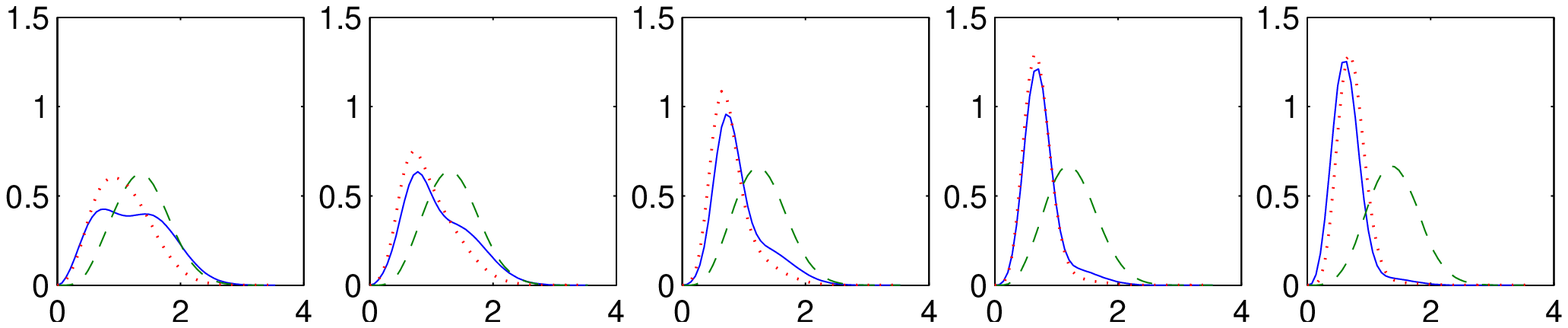}}
\subfigure[Normalized probability density vs. $\theta$]
        {\includegraphics[width=2\columnwidth]
        {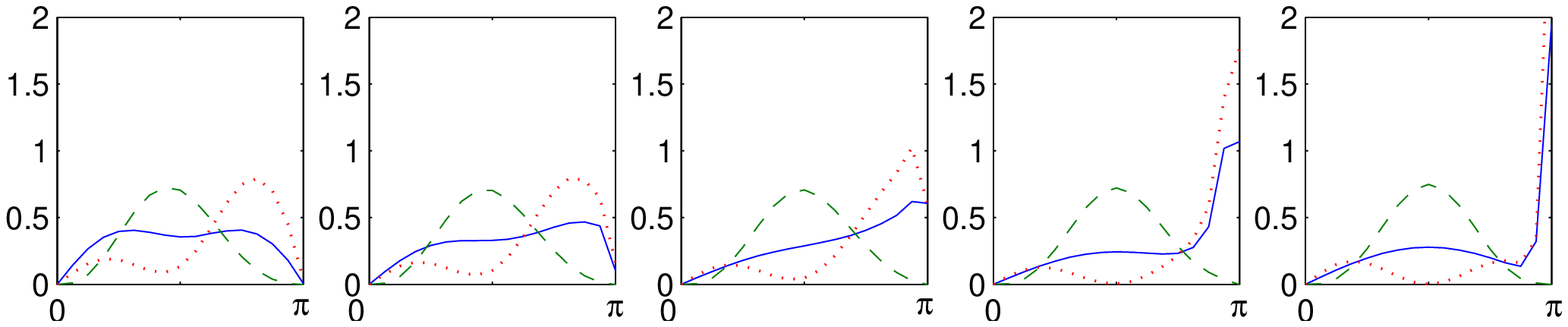}}
\caption{(Color online) Normalized probability density distribution
functions of variables defined in Fig. \ref{fig:vars} for various
scattering lengths. The solid line is for the $n=0$, $l=0$, $m=0$
state; the dashed line is for the $n=0$, $l=1$, $m=\pm 1$ state; and
the dotted line is for the $n=0$, $l=1$, $m=0$ state.}
\label{fig:dists}
\end{figure*}

We anticipate that the low-energy physics should be contained in a
truncated Hilbert space containing only the lowest few asymptotic
atom-dimer energy levels. Then Eq. \eqref{eq:eigen} is easily solved
numerically. We have checked that, indeed, the solution for the
ground state and the first excited manifold become insensitive to
the cutoff, as long as the first four or five atom-dimer energy
levels are included. For all results presented in this paper, we
have kept the first five energy levels. Because of the degeneracy of
the excited levels, Eq. \eqref{eq:eigen} becomes a $35\times 35 $
matrix equation. For a given energy, we solve numerically to obtain
the corresponding scattering length and eigenstate. By sweeping
through a range of energies, we map out the spectrum shown in Fig.
\ref{fig:Evsa}. At unitarity, our result for the low-energy spectrum
agrees with the analytic result of Ref. \cite{Werner06}. Upon
careful inspection, one can discern the presence of level crossings.
More usefully, in Fig. \ref{fig:Ediff} we display the difference
between the energy of three fermions in a single lattice site and
the energy of two fermions ($\uparrow \downarrow $) in the site and
the extra fermion alone in a separate site. Here we have used the
well-known exact solution for the two-body energy $E_{2}$
\cite{Busch98},
\begin{equation}
\frac{d}{a}=2\frac{\Gamma \left( \frac{3/2-E_{2}/\hbar \omega
}{2}\right) }{ \Gamma \left( \frac{1/2-E_{2}/\hbar \omega
}{2}\right) }
\end{equation}

Figure \ref{fig:Ediff} is our main result. There are two main
features we would like to point out. First, it is clear that it is
energetically favorable for atoms in a lattice to arrange themselves
such that there are less than three atoms per site, regardless of
the scattering length. This has already been assumed in the
derivation of an effective many-body Hamiltonian for atoms in an
optical lattice across a Feshbach resonance \cite {Duan05}, and is
confirmed by Fig. \ref{fig:Ediff}. Second, the level crossing in the
ground state is now quite evident. On the positive scattering length
side of the crossing (the BEC side, where the many-body system forms
a Bose-Einstein condensate of bound dimers), the ground state is
nondegenerate. On the other side (the BCS side, where the many-body
system forms a Bardeen-Cooper-Schrieffer superfluid of atomic Cooper
pairs), it is triply degenerate. Other crossings appear in the
excited spectrum, although to obtain quantitatively accurate results
for these one should include higher modes when solving Eq.
\eqref{eq:eigen}.

The origin of the level crossing is the differing symmetries of
the eigenstates. For very small, positive scattering length (deep
BEC side), formation of tightly bound dimers is favorable, so the
state should behave as the ground state of the relative atom-dimer
motion, which has s-wave symmetry. For very small, negative
scattering length (deep BCS side), the atoms are essentially
non-interacting, so the ground state comprises two atoms
($\uparrow \downarrow $) in the ground state of the trap plus the
third in the first excited state of the trap (which is triply
degenerate) because of Pauli exclusion. So, on the deep BCS side,
the ground state has p-wave symmetry. Due to the rotational
symmetry of a spherical harmonic trap, the total angular momentum
of the three particles should be a conserved quantity. However,
from the above analysis, this quantity has different values for
the ground state in the deep BEC and deep BCS limits. Therefore,
there must be a ground-state level crossing for this system as one
scans the scattering length. If one considers multiple lattice
sites with each site having on average two spin $\uparrow $ and
one spin $\downarrow $ atoms (which could be realized with
polarized fermions in an optical lattice with appropriate filling
number and population imbalance), as the three-body problem has a
level crossing with different ground state degeneracies in the BCS
and the BEC limits, there could be a corresponding quantum phase
transition for this many-body system (with small tunneling between
lattice sites) as one scans the scattering length.

The wavefunction given in Eq, \eqref{eq:SE3} does not generally have
definite \textit{relative} angular momentum for any two fermions. However,
in the limit as the distance between two distinguishable fermions goes to
zero, the wavefunction takes on the symmetry of the asymptotic atom-dimer
wavefunction in the remaining coordinates. This is a relative angular
momentum eigenstate due to the spherical symmetry of the limiting case. With
the relative coordinates defined by Fig. \ref{fig:vars}, the symmetry of the
wavefunction as a function of $R$ and $\theta$ in the limit as $r$ goes to
zero is shown by Fig. \ref{fig:sym}. For finite $r$, the $m=0$ wavefunctions
are affected by the asymmetry and take nontrivial shapes. We have plotted an
example in Fig. \ref{fig:asym}. Note that in this figure we include a factor
of $\sin \theta$ since the wavefunction itself diverges at $R=r/2$, $\theta
= \pi$ according to the boundary conditions we have imposed. On the BEC
side, the lobes are tightly bunched near $\theta = \pi$, but on the BCS side
they spread around as expected.

In general, the eigenstates are rather difficult to visualize, since
they depend nontrivially on three spatial variables as well as the
scattering length. However, one can get some idea of the evolution
of the eigenstates from Fig. \ref{fig:dists}, which shows the
normalized probability density as a function of the variables
introduced in Fig. \ref{fig:vars} for various scattering lengths. In
each subplot we have numerically integrated over the other two
variables to obtain a one-dimensional function. The dimer size, $r$,
clearly decreases in size as one enters the BEC regime, developing a
strong peak at the origin. However, as is clear from Fig.
\ref{fig:vars}, there are two ways to form a tightly bound dimer
(due to the two identical spin $\uparrow $ atoms) and we have
arbitrarily chosen one to define the origin $r=0$. So it is not
surprising that we see a more diffuse second peak at large distance
$r$, corresponding to the dimer forming between the spin $\downarrow
$ and the other spin $\uparrow $ atom. This is also the meaning of
the spike at $\theta = \pi$ on the BEC side.
\section{Summary}
We have found the low-lying energy levels of three fermions in a
harmonic trap and examined the corresponding wavefunctions. The
ground state has s-wave symmetry on the BEC side of Feshbach
resonance and has p-wave symmetry on the BCS side. In the resonance
region there is a level crossing, which may indicate a phase
transition in the corresponding many-body case. We also note that,
in the vicinity of resonance, the energy of three atoms in a single
site is greater than their energy if they are in two sites, with a
gap on the order of the trap spacing, validating the approximation
in Ref \cite{Duan05}.

This work was supported by the MURI, the DARPA, the NSF award
(0431476), the DTO under ARO contracts, and the A. P. Sloan
Fellowship.

Note added: After completion of this work, we became aware of a
recent work \cite{Stetcu07} which treats the trapped three-fermion
problem with a different approximation method.

\end{document}